# Where onshore wind meets forests: electricity system planning with ecologically graded forests

Muhammad Shahzad Javed[1,*], Marianne Zeyringer[1], Johannes Rahlf[2], Oliver Moen Snoksrud[2]

[1]Department of Technology Systems, University of Oslo, Norway.

[2]Norwegian Institute of Bioeconomy Research (NIBIO), Høgskoleveien 7, Ås, 1433, Norway.

[*]Corresponding author(s). E-mail(s): m.s.javed@its.uio.no

## Graphical abstract

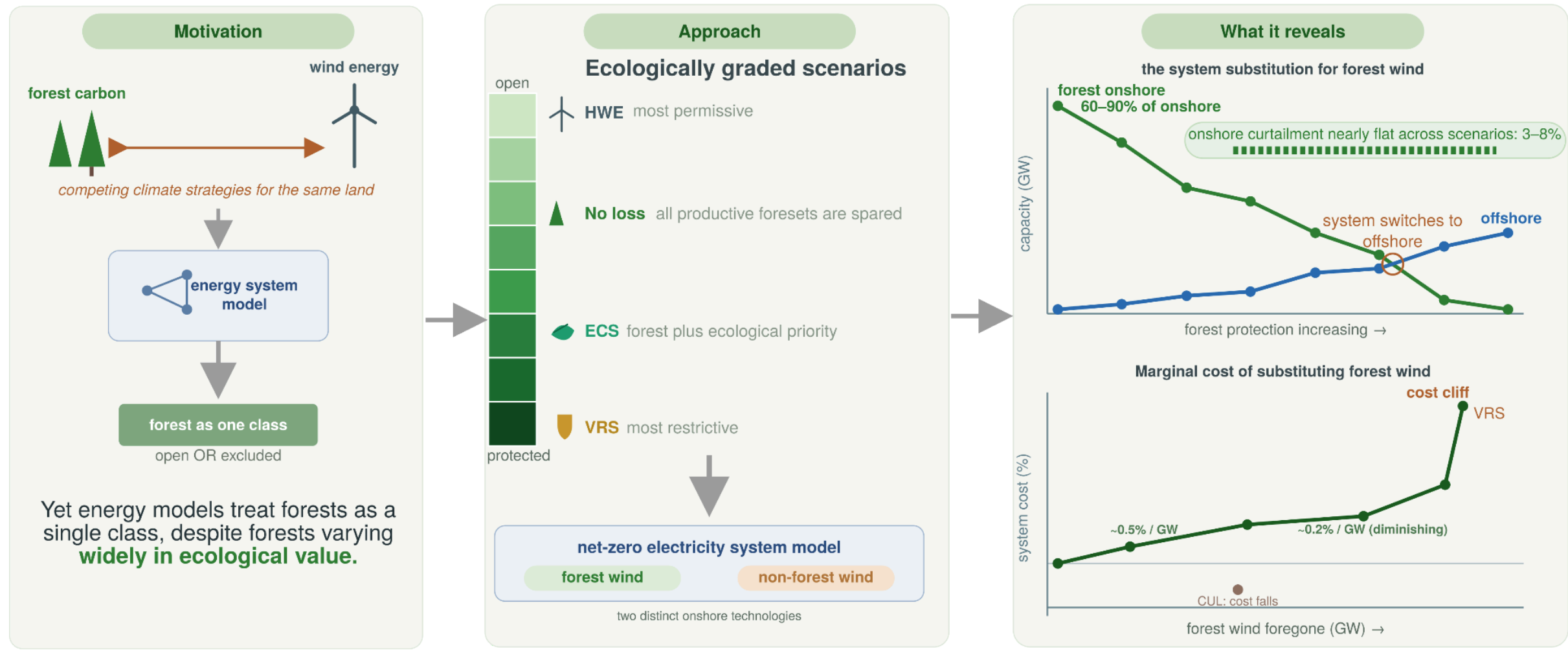

## Abstract

Onshore wind is expected to supply much of the growth in renewables, yet its expansion increasingly extends into forests, where wind generation and forest conservation act as competing climate strategies for the same land. Net-zero system assessments typically treat forests homogeneously. Here, we grade forest exclusions by ecological value and couple them with a high-resolution net-zero electricity system model, representing onshore wind on forested and non-forested land as two distinct technologies. Applying a set of ecologically graded scenarios to Norway, we quantify how forest protection by its ecological value reshapes system design, the balance of onshore wind within and outside forests, and the robustness of these outcomes to complementary energy policy levers. At equal cost, the system treats forest as its preferred onshore resource, with forest wind comprising 60–90% of onshore capacity under permissive scenarios. Excluding forest raises system cost by 0.16–0.5% per GW of foregone forest wind, with diminishing returns as ecological criteria for forest exclusions tighten. Directing wind into forests does not compromise system efficiency, with onshore curtailment remaining largely unchanged at 3–8%. Land-use and technology decisions that appear independent may, in fact, interact, as solar deployment reshapes the value of each wind resource and determines which is displaced from the generation mix. By quantifying the ecological value forgone per unit of forest onshore wind, this study makes explicit a trade-off that energy system modelling usually leaves implicit, providing a transferable basis for balancing the competing climate strategies of protecting forests and sustaining wind expansion.

**Keywords:** onshore wind siting; ecologically graded forests; biodiversity conservation; electricity system optimisation; net-zero power systems

## 1. Introduction

Achieving global decarbonisation targets increasingly relies on rapid electrification, which is expected to drive a substantial expansion of variable renewable capacity [1,2]. Onshore wind is anticipated to supply much of this growth [3], as one of the most mature and cost-competitive low-carbon technologies currently available [4]. How much of its potential can ultimately be realised, however, may depend less on cost than on the land that societies are willing to allocate to it. In this sense, the energy transition is increasingly a question of land allocation as much as of energy supply [5]. It is particularly exposed to this constraint, because generating at scale requires wind turbines to be distributed across large areas rather than concentrated on a compact footprint [6]. As comparatively uncontested sites become scarce [7,8], deployment tends to expand into areas that societies value for other purposes, including outdoor recreation, cultural landscapes, and biodiversity conservation [9]. This expansion is leading to widespread public opposition, shaped by a recognisable set of concerns, including the visual impact of turbines on scenic landscapes [10,11], noise and shadow flicker [12], and changes in biodiversity arising from habitat loss [13,14]. Because decisions on where wind projects are built increasingly rest with local communities and municipal authorities [15], while their consequences for the energy system materialise at the national scale, these constraints can propagate upward to influence system-wide costs, the technology mix, and ultimately the feasibility of the transition [16,17].

Forests have become an increasingly prominent frontier for this expansion, with a growing share of new onshore wind capacity deployed on forested land, particularly across Northern Europe [18,19]. Such locations can attract development, offering large, sparsely settled areas where land is comparatively inexpensive and opposition tends to be weaker than in open or inhabited landscapes [20]. Forests are not straightforward hosts, however, because installing turbines requires clearing canopy to build related infrastructure, which can make wind development a direct and localised driver of deforestation and fragmentation [18,21]. For example, forest clearance in Finnish wind-project areas averaged around six times that of comparable undeveloped land [21]; onshore wind in Norway has not avoided high-value habitats contributing to fragementation in forested landscapes [22], and in Scotland around three-quarters of targeted land comprised carbon-rich, semi-natural habitats whose conversion risks fragmentation and deforestation [23]. The consequences can extend beyond the loss of trees to broader ecological disruption, including heightened collision risk for bats that aggregate at forest gaps, disturbance of birds, and reductions in downwind vegetation productivity [24–29]. These impacts are especially consequential because forests are among the most important terrestrial stores of carbon and biodiversity, yet in many regions the forests now hosting wind are not untouched wilderness but long-managed commercial production forests shaped by generations of timber harvesting [30] **[ref]**.

Framing forests only as areas to protect from onshore wind, however, obscures a deeper tension, since maintaining forest carbon and expanding onshore wind are competing climate solutions [31,32]. The strength of this trade-off is likely to vary with region and forest type, depending both on the carbon and biodiversity a forest holds, which is shaped by attributes such as age and biomass [30], and on how strongly forest cover suppresses onshore wind potential [31]. Excluding wind from forests is therefore not without cost, since it forgoes low-carbon generation and can displace development onto other valued land or to costlier alternatives. Slower decarbonisation carries its own ecological cost, since unmitigated climate change is itself a growing driver of biodiversity loss [32]. In Norway, for instance, only about 1.7% of forests are old forests (> 160 years)[1], so excluding a given forest may safeguard very different amounts of biodiversity, and steering wind toward lower-value forest could partly ease the trade-off between climate action and biodiversity [33]. The question is therefore not whether wind belongs inside or outside forests, but how to deploy it so that decarbonisation advances while forests of higher ecological value are spared, a distinction existing assessments have largely overlooked.

[1] https://tv.nrk.no/serie/oppsynsmannen

A growing body of research has sought to reconcile wind expansion with competing environmental and social values, yet across its main domains forests tend to be represented homogeneously. Policy-oriented studies frequently identify forests as carbon stores and biodiversity-rich areas to be shielded from wind development, without accounting for the ecological variation among them [30,34,35]. Spatially explicit siting studies address such concerns by mapping suitable land based on, among others, bird sensitivity, scenicness, landscape character, and protected status [5,7,10,36–38]. Within these analyses, forests are typically treated as a uniform land class, so that variation among forests is not explicitly considered. Energy system models have increasingly quantified how social and environmental restrictions reshape future energy system designs by explicitly treating forests as a binary class (forest or non-forest) [16,17,39,40]. Modelling that incorporates stakeholder preferences has enhanced the realism of this work, yet it commonly treats forests as a single category [41–44]. Across these domains, the ecological differentiation of forests, when considered alongside other environmental criteria, has not been linked to energy system modelling outcomes, leaving open how ecologically graded, spatially explicit forest classes would reshape the cost, technology mix, and spatial deployment of a net-zero electricity system.

Here we address **this gap** by linking ecologically graded, spatially explicit forest classes, considered alongside other environmental criteria, to a model of a future net-zero electricity system for Norway. Within this framework, onshore wind on forested and non-forested land is represented as two distinct resources, so that the system can reveal where wind is drawn as forest protection varies. This paper answers the following research questions. ***First***, how do ecologically graded forest choices reshape onshore wind deployment in the net-zero power system designs? ***Second***, how robust are these outcomes to policy choices on solar capacity, transmission expansion, and buffers around forests? ***Finally***, how does the system trade off onshore wind across forested and non-forested landscapes under ecologically varying forest-siting choices?

## 2. Methods

The tensions that motivate this study are especially acute in Norway. Rapid electrification is expected to raise electricity demand from around 127 TWh in 2023 to 250 TWh by 2050 [45], with studies indicating that supply may fall short of demand before 2030 if no additional generation capacity is added to the system [16]. Large-scale hydropower, the backbone of the existing system, has been largely foreclosed for further expansion since 2001[2], leaving onshore wind as the principal scalable domestic resource. Yet onshore wind grew from 1.7% to 10% of generation between 2017 and 2022 before stalling after 2020 [46], as public opposition intensified around the 2019 national framework for wind siting [15]. Norway has strong onshore wind resources [7] but has installed far less capacity than Nordic neighbors such as Sweden[3], partly reflecting resistance to development in valued landscapes [43,47]. It is also among Europe's more forested countries, with forests covering roughly 38% of its land area and productive forest managed for timber making up most of it[4]. Because wind energy is likely to compete with forested land, Norway provides an ideal case for assessing how ecologically differentiated forest protection reshapes a net-zero electricity system, with findings relevant to other countries where forest carbon and expanding onshore wind emerge as competing climate solutions.

### 2.1. Scenarios

[2] https://www.nrk.no/urix/vannkraftepoken-er-slutt-1.462300
[3] https://ourworldindata.org/grapher/cumulative-installed-wind-energy-capacity-gigawatts
[4] https://www.regjeringen.no/en/topics/food-fisheries-and-agriculture/skogbruk/innsikt/skogbruk/id2009516/

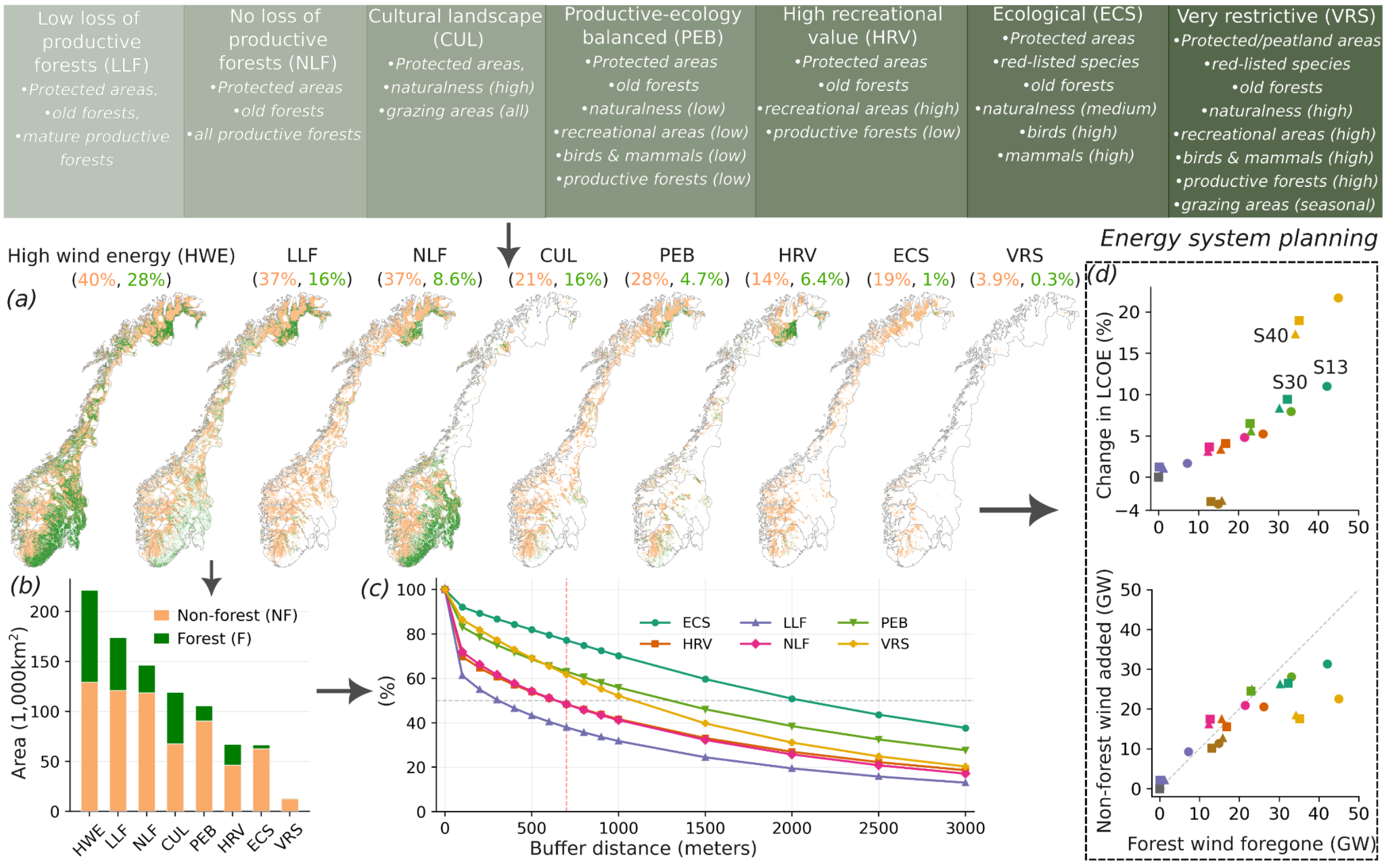

**Fig. 1.** Ecologically graded forest exclusion scenarios and their impacts. The scenarios span a gradient from the most permissive (HWE) to the most restrictive (VRS); the detailed criteria exclusions are summarised in Table A1. **(a)** Land available for onshore wind development. Percentages illustrate the share of the country's total forest area (green) and total non-forest area (orange) that remains available; for example, under HWE 40% of non-forest is available. **(b)** Total available area under in 1,000 km². **(c)** Effect of applying a protective buffer around ecologically graded forest exclusions (see section 2.2). The red vertical line marks the 700 m buffer, and the horizontal line marks the 50% reduction in its initially available land at zero buffer. **(d)** System-level design impacts, with changes in system-levelised cost of energy (LCOE) and substitution patterns for each forest's onshore wind foregone (Fig. 2 & Fig. 4). Marker color indicates the scenario, and marker shape indicates solar growth (Table 1).

No single exclusion rule can capture how forests should be weighed against wind, because forests differ widely in ecological value and rarely stand apart from other contested land uses [30,36]. We therefore construct eight scenarios that grade forests by ecological value and combine this grading with other environmental criteria, spanning a gradient from permissive to restrictive development [8]. Rather than uniformly excluding all protected areas, we vary protected-area exclusions by scenario, informed by an overlap analysis of Norway's existing wind areas against the Common Database on Designated Areas (CDDA), a European inventory of officially protected areas [17]. That analysis indicates that past siting has not consistently avoided even strictly protected land, with some existing wind areas overlapping CDDA category Ia areas (strictest protection)[5]. Each scenario therefore excludes only the protected-area categories relevant to its priorities; the very restrictive scenario excludes all categories **(Table A1)**. The most permissive scenario, ***high wind energy (HWE)***, maximises deployment by opening nearly all forest to development **(Fig. 1)**. The ***cultural scenario (CUL)*** protects areas of high naturalness together with all reindeer grazing areas and migratory routes, reflecting the cultural landscapes of Sámi reindeer husbandry, which are concentrated in northern Norway, and thereby redirects development toward productive forest nearer the southern demand centres. Two further scenarios prioritise forestry production to differing degrees. ***Low loss of forest production (LLF)*** permits wind only in forest of low or very high timber volume, that is young or near-harvest forest, while excluding forest of intermediate productivity and forest older than about a century. ***No loss of forest production (NLF)*** tightens this further by excluding all productive and high-biomass forest, so that development is confined to non-productive, low-biomass forest.

[5] https://wimby.eu/resource/d1-3-land-and-sea-use-and-change-maps/

The remaining scenarios layer broader ecological and social criteria onto forest protection. ***Productive-ecology balanced (PEB)*** excludes forests that are at once highly productive and ecologically valuable, combining thresholds on productivity, biomass, and naturalness with bird- and mammal-sensitive areas and high-use recreation areas. ***High recreational value (HRV)*** instead centres on forests valued for recreation and landscape, excluding buffered recreation areas together with productive and high-biomass forest [15]. The ***ecological scenario (ECS)*** prioritises biodiversity by excluding old forest, high-naturalness areas, red-listed species habitat, and bird-sensitive sites, yet deliberately leaves productive forest open so that development is steered toward already-managed, productive forest. The very ***restrictive scenario (VRS)*** represents maximal protection, adding to these ecological exclusions all productive and high-biomass forest, recreation and reindeer areas, peatlands, and the strictest naturalness threshold, so that only about 0.3% of forest land remains available **(Fig. 1)**. The land available for development under each scenario, within and outside forests and across buffer distances, is summarised in Fig. 1, and the underlying datasets and spatial processing are described in Section 2.4.

The thresholds that separate excluded from available land were set differently for the ecological and the forest datasets, reflecting how each is best interpreted. For the continuous ecological surfaces, namely bird and mammal sensitivity, recreation intensity, and naturalness, we applied Jenks natural-breaks classification to partition areas into five classes that maximise within-class similarity and between-class contrast [48]. Natural breaks were preferred over fixed cut-offs because they follow the actual distribution of each surface rather than imposing arbitrary boundaries. These classes were then read through the lens of Environmental Impact Assessment (EIA) frameworks [49], which grade environmental value from negligible to very high and treat the upper classes as the most sensitive to disturbance, thereby progressively excluding the higher-value classes under stricter scenarios. The forest datasets, by contrast, were graded by their defining attributes of age and biomass, with thresholds informed by forestry expertise and Norwegian forest policy frameworks rather than by statistical classification. Buffer distances around excluded forest follow the same logic developed in **Section 2.2**.

These scenarios are not forecasts of policy but analytical constructs for tracing how a net-zero system responds when forest carbon and onshore wind, two climate strategies competing for the same land, are reconciled in different ways [31,32]. By making the ecological grading of forests explicit rather than encoding it implicitly within an optimisation objective, this approach reveals the system-level consequences of these choices while leaving the preferences themselves to society, a perspective still rarely taken in energy system modelling [16,17,40,50–52].

## 2.2 Policy levers in scenarios

To capture uncertainty in complementary technology pathways, we vary two system-level policy levers alongside the exclusion scenarios described above: solar deployment capacity and transmission network expansion.

Norway's solar PV sector is in a formative phase, contributing less than 1% of total electricity generation as of 2024 [17]. NVE projects solar electricity production to increase from 0.45 TWh in 2023–2024 to 9 TWh by 2050, roughly corresponding to 12-13 GW under Norwegian solar conditions[6]. However, estimates of Norway's long-term solar potential diverge substantially from 3.2 GW by 2030 based on historical growth rates [16] to 24 TWh/year from building-applied potential alone [53], and up to 63 GW by 2050 under accelerated growth assumptions [17]. To span this range, we model four solar capacity constraints: 13 GW (S13) (aligned with NVE's central projection), 30 GW (S30), 40 GW (S40), and unconstrained (S+) (**Table 1**). Moreover, other factors such as grid integration, battery costs, material scarcity, or financial profitability may also affect solar development. This allows us to assess how onshore wind exclusion impacts interact with solar deployment pathways, given that solar PV increasingly substitutes for restricted onshore wind sites due to its low cost and proximity to southern and eastern demand centers.

[6] Norges vassdrags- og energidirektorat, Utviklingen i Kraftmarkedet Mot 2050

For transmission, we model two configurations: fixed transmission (Trans) at existing inter-regional (county) capacities and cost-optimized transmission expansion (Trans+), in which the model can invest in new inter-regional links. Norway's transmission development trajectory remains contested, with ongoing policy debates about both internal grid reinforcement and cross-border interconnection capacity[7]. We hold Norway's existing import/export capacity constant across all configurations and let the model optimize dispatch under this constraint. Combined with the four solar growth assumptions, this yields eight system configurations per exclusion scenario, enabling a systematic decomposition of how technology policy choices interact with onshore land-exclusion preferences.

Buffer distances around excluded forest areas constitute the third policy lever, applied within each exclusion scenario. Regulatory setback requirements for onshore wind vary widely across Europe depending on societal tolerance [17], infrastructure type [54], landscape scenicness [10], ecological sensitivity [14], and other environmental factors. Empirical studies document measurable ecological impacts, including reduced vegetation productivity and altered wildlife behavior extending up to 1000m from wind turbine locations [28], with species activity reduced by 20-54% [26], especially elevated collision risk of bats and birds within a 1000m radius of forest edges [24].

On the other hand, the literature indicates that more restrictive setback distances can substantially increase the social costs of wind energy, though the magnitude is highly site-dependent, and differentiated setbacks (varying by settlement type or ecological sensitivity) achieve substantially lower costs than uniform distance requirements [55]. Forest exclusion zones and related setback distances create opportunity costs that depend on the spatial correlation between excluded areas and nearby high-quality wind resources, a trade-off that is particularly relevant to Norway, where forest exclusions overlap with productive wind sites in southern and Eastern Norway. In the absence of established regulations on setback distances for wind turbines near high biodiversity areas [13], we vary buffer distances from 0 to 3000m around excluded forests in 100m increments up to 1000m, then in 500m increments to 3000m. This range encompasses current Norwegian practice (~800m from settlements [56]), precautionary distances supported by ecological evidence

[7] https://www.statnett.no/for-aktorer-i-kraftbransjen/planer-for-stromnettet/omradeplaner/

(~1000m for bat and bird activity zones), and extreme distances (2000–3000m) representing potential future tightening of environmental regulations.

**Table 1.** System configurations based on solar deployment capacity, transmission expansion, and buffer distances around excluded forest areas. Solar capacity represents the maximum new solar PV installable by 2050. Buffer distances are applied to the six scenarios with forest-based exclusions (LLF, NLF, PEB, HRV, ECS, VRS). Each configuration is combined with all eight exclusion scenarios, yielding 736 unique model runs.

| **Configuration** | S13 | S13+ | S30 | S30+ | S40 | S40+ | S+ | S++ |
|---|---|---|---|---|---|---|---|---|
| **Solar capacity [GW]** | ≤13 | ≤13 | ≤30 | ≤30 | ≤40 | ≤40 | OPT | OPT |
| **Transmission** | Fixed | Expand | Fixed | Expand | Fixed | Expand | Fixed | Expand |
| **Buffer to excluded forests (m)** | 0, 100, 200, 300, 400, 500, 600, 700, 800, 900, 1000, 1500, 2000, 2500, 3000 | | | | | | | |

**2.3 Modelling setup**

We model the Norwegian electricity system using a modified version of highRES, an open-source, high-resolution capacity-expansion model that co-optimizes generation, transmission, and storage investments together with hourly operation under a least-cost objective and perfect foresight [57,58]. Its adaptation to Norway, together with the full formulation and techno-economic parameters, can be found in the authors' previous studies [8,43]. Electricity demand for 2050 is taken from the transmission system operator´s (Statnett) projections of about 250 TWh [45], which do not yet reflect deep-decarbonisation pathways and may therefore understate demand by a further 30 to 50 TWh due to hydrogen and synthetic-fuel production. Existing hydropower and transmission are retained and held fixed; existing onshore wind of around 5 GW is included, with its footprint removed from developable land; and we adopt a conservative capacity density of 2.4 MW/km² [17] rather than the 8.6 MW/km² projected by NVE for Norwegian landscapes [10,59]. We do not consider existing offshore wind capacity because it is either used as a testing facility or operates off-grid, powering oil and gas fields in Norway. All scenarios share identical demand, weather, and techno-economic assumptions and the solar, transmission, and buffer levers defined in Section 2.2, so that the sole cross-scenario variation is the forest and environmental land exclusions.

Three features distinguish modelling setup of this study. First, onshore wind is disaggregated into forest and non-forest resources as two distinct technologies with identical techno-economics, classified from CORINE land cover, so that the systems' substitution between them is revealed directly as forest exclusions vary. Second, two wind-speed concerns are addressed separately, both of which are particularly relevant to a forest-focused study. Because Norway's complex terrain leaves the 0.25° ERA5 reanalysis unable to resolve localised wind, capacity factors are bias-corrected for each grid cell against observed NVE wind-power production [16,43]; and because canopy roughness could in principle lower the forest wind resource, we confirmed that the ERA5 land representation already captures this wake effect, in broad agreement with the CORINE forest classification, so no additional capacity-factor penalty is applied to forest sites. Third, distance-based offshore grid-connection costs are added following [17]. Scenario exclusions are applied at 1×1 km resolution and aggregated to county level for optimisation, using the 2010 weather year, with data sources and processing detailed in **Section 2.4**.

### 2.4 Geospatial data and processing

The scenario exclusion criteria draw on high-resolution spatial datasets, listed with their sources and formats in **Table A2**. Biodiversity is represented by two components. Species richness for 247 bird and 5 bat species was derived from validated life-cycle-assessment distributions built from Global Biodiversity Information Facility (GBIF) occurrence records and habitat-suitability modelling [22,60]. Red-listed species occurrences were obtained from the Norwegian Red List for Species, which classifies species into IUCN-based extinction risk categories. Forest ecological value is captured through three layers, namely a map of forest established before 1940 and not subsequently clear-cut, a modelled probability of natural forest, and a continuous naturalness gradient reflecting dead-wood volume, dead-wood variation, and forest development stage, together with forest productivity, standing biomass, and age. Protected areas were taken from a national protected-areas register and classified by IUCN management category, while peatlands, water bodies, urban areas, and the forest and non-forest distinction were derived from CORINE land cover. Recreational areas and reindeer seasonal grazing areas with migratory routes were obtained from national mapped recreation and reindeer-husbandry datasets. All datasets were reprojected to the ETRS89-LAEA coordinate reference system (EPSG:3035).

For each scenario, the relevant exclusion layers were combined into a composite availability mask separating available from unavailable land at 1 × 1 km resolution. The very restrictive scenario, for instance, excludes all protected categories, red-listed species locations, peatlands, the oldest and most natural forest, recreation and reindeer areas, and all productive and high-biomass forest, whereas the high wind energy scenario excludes only strictly protected areas. Each mask was then intersected with the CORINE forest classification to distinguish available and unavailable land within and outside the forest (**Fig. 1b**), yielding the two onshore wind resource datasets used in highRES. These scenario-specific availability rasters were then combined with default technical and environmental exclusions applied uniformly across all technologies, including slope, elevation, physical infrastructure, urban areas, and water bodies, with buffer distances set according to the type of infrastructure [8,16,17,43]. This spatial union operation across multiple raster layers was performed using the 'atlite' geospatial toolbox (version 0.2.13) [61]. Availability rasters for these areas were aggregated to a 30 × 30 km resolution and used as input, along with ERA5 weather data, to optimize capacity expansion in highRES (**Section 2.3**).

# 3. Results

## 3.1 System cost impacts

Across the most permissive solar and transmission assumptions, Norway's net-zero electricity system achieves an LCOE of ~603 $NOK_{2025}$/MWh, rising to ~634 $NOK_{2025}$/MWh when solar capacity is constrained to 13 GW and transmission is fixed at current levels (**Fig. 2a)**. This ~5% spread indicates that the cost impact of technology and infrastructure assumptions alone is comparable in magnitude to that of several exclusion scenarios (2–7% at zero buffer, **Fig. 2b)**, suggesting that future solar deployment rates are as consequential as land-use choices for net-zero electricity costs.

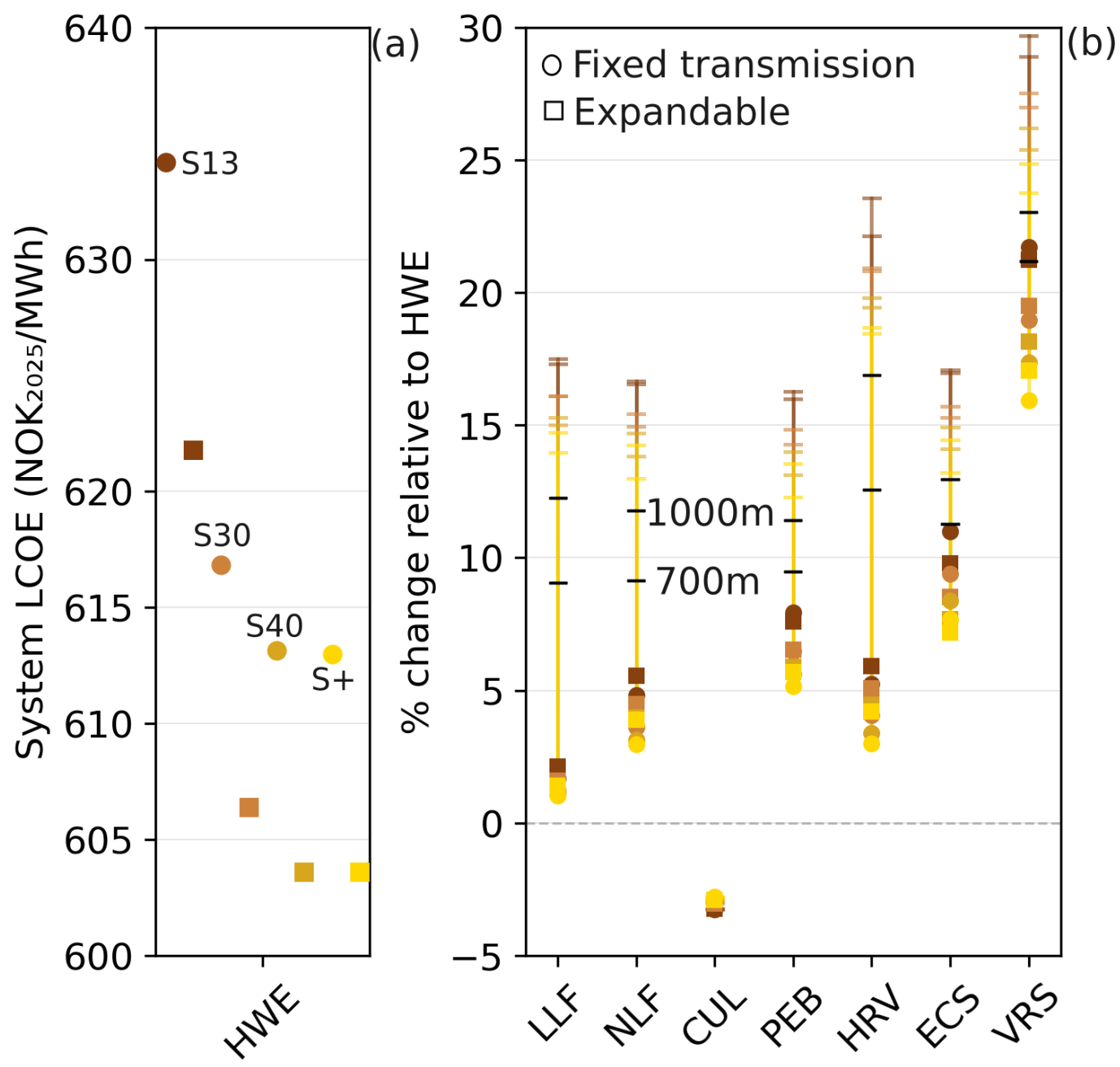

**Fig. 2.** System levelised cost of energy (LCOE) for a 2050 net-zero national electricity system across scenarios and system configurations. (a) Absolute LCOE under the most permissive scenario (HWE) across varying solar (e.g., S13 = 13 GW solar capacity constraint) and transmission assumptions. (b) Percentage change in LCOE relative to each configuration's own HWE baseline. Colored dots indicate zero-buffer values; vertical lines extend upward to the maximum cost at a 3000 m buffer around excluded forests, illustrating the additional cost of the buffer policy. Black horizontal markers indicate the mean cost increase across configurations at buffer distances of 700m and 1000m. Darker markers correspond to more constrained solar assumptions **(see Table 1)**.

The cost of moving from the least restrictive scenario (HWE) to progressively stricter exclusion criteria follows a non-linear gradient **(Fig. 2b)**. Scenarios that primarily restrict forest access while preserving non-forest wind potential — LLF and NLF — increase system costs by 2–5% at zero buffer distance, indicating that the system can largely absorb the loss of forest-based wind sites through substitution with other available resources [8,16]. CUL criteria achieve negative cost increases of approximately -3% because related exclusions remove almost all of northern Norway and redirect new onshore deployments towards available productive wind sites closer to electricity demand centers, reducing system bottlenecks such as transmission congestion [17].

Scenarios imposing broader ecological and social restrictions — PEB, HRV, and ECS — incur 5–10% cost increases at zero buffer, reflecting the progressive elimination of cost-effective onshore wind sites, both within forests and outside them. The most restrictive scenario (VRS) raises costs by 16–22% with a zero buffer across configurations, widening the spread between configurations to ~6 percentage points, indicating that the value of solar flexibility and transmission expansion increases as onshore wind constraints tighten [16].

A buffer around excluded forest areas substantially amplifies these costs, but with diminishing marginal impact [55,62]. The 700-1000 m regulatory milestones capture approximately 50–60% of the total buffer cost range (0–3000m) [56]. Beyond 1000m, each additional meter of buffer contributes progressively less to system cost. At the extreme (3000m buffer), LLF and NLF reach 15–17% cost increase, PEB reaches 14–16%, HRV reaches 18–24%, ECS reaches 16–18%, and VRS approaches 22–29%, exceeding the combined cost mitigation (~5%, **Fig. 2a)** achievable through solar expansion, transmission investment, and offshore wind substitution, indicating that cost increases due to lost onshore capacity at extreme buffer distances cannot be fully compensated with any combination of available technology alternatives.

Notably, the vertical ordering of dots within each scenario is consistent: darker markers (solar-constrained configurations) always sit above lighter ones (solar-unconstrained), indicating that solar deployment is a robust cost-reduction lever that partly offsets the loss of productive onshore land sites, regardless of which exclusion criteria are applied. It is worth noting that the cost gap between configurations (coloured dots) remains limited to 3–6% even under VRS, despite the increased proportional solar saturation up to 45 GW (**Fig. 3**). This indicates diminishing returns from solar expansion [63]: the system still requires onshore wind capacity for periods when solar cannot complement or serve (e.g., peak evening times or winter) [8], and the widening temporal gap between solar and wind results in costs rising by up to 29% above HWE, regardless of solar deployment level.

Transmission expansion (square vs. circle markers) yields a 0–1% cost reduction, with its effect most visible under solar-constrained configurations, where spatial redistribution of wind generation has a higher marginal value. Under solar-unconstrained configurations, transmission provides minimal additional benefit (except VRS), suggesting that transmission expansion acts as a partial substitute for local generation adequacy rather than a complementary investment to solar.

### 3.2 Impact on national technology mix

**Fig. 3a** shows the base case technology composition (S13) across all eight scenarios. Under HWE, onshore wind dominates at approximately 56 GW (~10 GW outside forest, ~46 GW in forest), with offshore wind contributing ~1 GW. The substantially higher forest wind deployment compared to non-forest reflects that the system preferentially selects southern forest sites that combine strong wind resources with proximity to demand centers (**Fig. 6**). As exclusion criteria tighten, total onshore wind declines progressively — to ~53 GW NLF, CUL, ~43 GW under PEB and HRV, ~35 GW under ECS, and ~15 GW under VRS — while offshore wind expands from ~1 GW to ~18 GW. This substitution is incomplete (except LLF, where total capacity increases by ~3 GW as the restriction of productive forests redirects deployment to low-wind non-forest sites): total new generation capacity drops from ~70 GW (HWE) to ~48 GW (VRS), a ~31% reduction. These capacity gaps between onshore losses and offshore gains are the structural drivers of the cost increases observed across scenarios **(Fig. 2)**.

Forest wind capacity follows a distinct pattern across scenarios. It contributes up to ~46 GW under HWE and ~38 GW under LLF, where forest land is broadly available; drops to ~25–30 GW under NLF and CUL, where restrictions exclude natural and productive forest sites; declines to ~12–19 GW under PEB and HRV; and falls to ~1–3 GW under ECS and VRS. The magnitude of forest wind under permissive scenarios—constituting 60–90% of total onshore capacity—indicates that when forest and non-forest sites compete at equal cost, the system strongly favors forest deployment, consistent with the empirical concentration of onshore wind on forested land across the Nordics [19,21]. Notably, HRV and PEB achieve similar total onshore capacity through different spatial strategies—HRV deploys more in forests (~19 GW), while PEB concentrates deployment outside forests with ~12 GW in forests.

The system cost of the foregoing forest wind is not a fixed value but depends on how forest restrictions interact with other exclusion preferences applied simultaneously. Between HWE (46 GW forest) and LLF (38 GW), a loss of ~8 GW of forest capacity increases costs by 2–3%, implying ~0.3–0.4% per GW—reflecting that the excluded productive forests contained economically attractive sites with limited substitutes at the margin. Between HWE and NLF, a further loss of ~13 GW of forest capacity raises costs to 4–5% above HWE, yielding ~0.19–0.24% per GW—a lower marginal rate that illustrates that the additional forest capacity lost carries comparatively lower substitution cost as the system still retains sufficient non-forest and offshore alternatives to absorb the reduction. Between NLF and ECS, where forest capacity drops from 25 GW to 3 GW, costs rise by a further 4–5%, implying ~0.16–0.23% per GW again. This declining marginal cost of forest exclusion—from ~0.4% per GW for the initial 8–10 GW foregone to ~0.2% for subsequent tranches—reflects the system's optimal site selection hierarchy: the most cost-effective forest locations are deployed first, and each successive tranche of excluded forest land displaces progressively less valuable sites that the system can substitute at lower marginal costs [64]. When coupled with increasingly restrictive non-forest land-use and technology-specific constraints (i.e., ECS and VRS; Fig. 2 & 3), the marginal cost per GW rises up to 0.5%, illustrating that the system's substitution capacity has a threshold, as the cost efficiency of spatial diversification erodes under very restrictive scenarios.

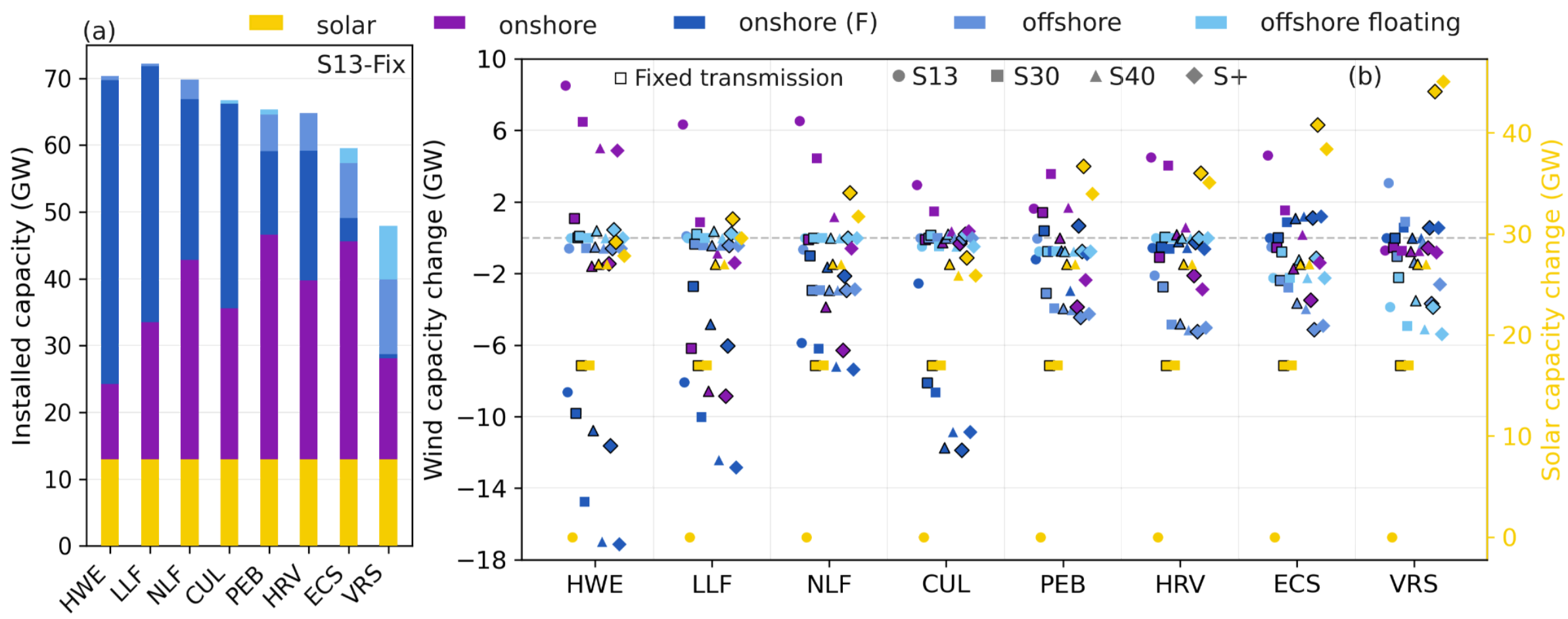

**Fig. 3.** System generation capacity for a 2050 national net-zero electricity system across scenarios and system configurations. (a) Technology composition under the base case configuration (S13), showing both onshore in the forests (onshore (F)) and outside forests (onshore). (b) Change in each technology's capacity relative to each scenario's own S13 base case. Dot color indicates technology; dot shape indicates solar configuration type; black border indicates fixed transmission. The right y-axis (yellow) shows solar capacity change. The dashed grey line marks zero change. See **Table 1** for configuration definitions and **Table A1** for scenario definitions.

### *3.2.1 Sensitivity to solar deployment and transmission expansion*

**Fig. 3b** shows how each scenario's technology mix responds to alternative configurations **(Table 1)** relative to the solar-constrained (S13) case. Three tiers of insights emerge.

First, transmission expansion alone produces scenario-dependent capacity shifts that are most pronounced under permissive scenarios. Under HWE and LLF, transmission expansion redistributes up to 6 GW between onshore forest and non-forest sites—reducing forest deployment while increasing non-forest capacity—as expanded inter-zonal links enable the system to access previously transmission-constrained remote non-forest sites [39]. Under more restrictive scenarios (PEB, HRV, ECS, VRS), transmission effects remain limited to 1–2

GW, primarily rebalancing between bottom-mounted and floating offshore. This asymmetry reflects that transmission expansion has greater leverage when the system has abundant deployment options to redistribute among, and diminishes when land constraints have already determined the generation portfolio. As evident from system cost results **(0–2%; Fig. 2)**, its effect scales with the breadth of available onshore alternatives.

Second, the effect of intermediate solar expansion (S30 and S40) on the technology mix is also scenario-dependent, revealing how exclusion preferences shape the system's substitution pathway. Across most scenarios from HWE through ECS, additional solar primarily displaces onshore forest wind by 2–14 GW, while non-forest onshore capacity often remains stable or increases by 2-8 GW, shifting to the positive side as the system retains non-forest sites that complement solar more effectively than the displaced forest sites [8,43]. In scenarios with the most restricted onshore availability (HRV, ECS, VRS), solar also displaces offshore wind by 2–4 GW, as total onshore capacity approaches its constraint floor and cannot be reduced further.

Third, unconstrained solar (S+) amplifies the aforementioned displacement patterns. Further solar saturation displaces onshore forest wind by 6–18 GW and non-forest onshore by 2–8 GW, with the forest-to-non-forest displacement ratio reflecting the relative abundance of each land type in each scenario's available capacity pool. The largest reductions in forest wind happen in HWE, LLF, and CUL scenarios, possibly due to its high saturation in the base solar-constrained scenario (S13), making it a primary substitution candidate for solar. In VRS, where onshore capacity is already minimal, solar displaces offshore wind by 4–6 GW, where offshore saturation was elevated to ~20 GW in the S13 base case. The consistent pattern across all scenarios is that solar preferentially displaces whichever wind technology occupies the largest share of the capacity mix—primarily forest wind in permissive scenarios followed by offshore wind in restrictive ones—reflecting temporal competition for dispatch value [63].

This solar-driven displacement of wind capacity carries a flexibility cost: battery storage requirements increase from 22–33 GWh under solar-constrained configurations to 33–48 GWh when solar is unconstrained **(Fig. A1)**, reflecting the need to absorb daytime solar surplus for evening and overnight dispatch [65]. Notably, within any given solar configuration, the difference in storage between the most permissive (HWE) and most restrictive (VRS) scenarios is 5–10 GWh, despite a fourfold difference in onshore wind capacity, likely illustrating that solar adoption levels are the primary driver of storage requirements, with land-use preferences playing a secondary role. In other words, the decision of how much solar to deploy determines how much storage Norway needs; the decision of which forests to protect does not.

In practical terms, each additional GW of solar serves a different purpose depending on the land availability preferences. The scenario-dependent substitution pattern demonstrates that land-use preferences made today determine not only how much onshore wind is built, but also which technologies solar will displace as it expands. The comparison between LLF and NLF isolates this mechanism. Under LLF, where productive forests remain partially available, unconstrained solar expansion displaces both forest wind (~12 GW) and non-forest wind (~6 GW) while offshore capacity remains essentially unchanged, mainly because sufficient forest capacity exists to serve as the adjustment variable. Under NLF, where all productive and natural forests are excluded, the same solar expansion replaces the onshore displacements with offshore (~4 GW) because the reduced onshore pool cannot fully absorb the solar substitutions and the displacement spills over into offshore capacity.

The combined effect of all configurations reveals that the vertical spread of capacity changes within each scenario spans up to 28 GW (from -18 to +10 GW for onshore wind), while horizontal differences between adjacent scenarios at the same configuration are typically 2–6 GW—reinforcing the cost analysis finding that technology deployment assumptions are as consequential as exclusion preferences for Norway's future electricity system.

### 3.3 Forest-specific onshore deployment analysis

The system's use of available land for onshore wind deployment reveals that forest areas are the preferred deployment resource **(Fig. 4a)**. The use of forest wind capacity potential (i.e., the share of available forest area deployed) ranges from ~25% (HWE) to ~78% (ECS), and consistently exceeds the non-forest capacity potential used across all scenarios except VRS. Non-forest capacity potential used ranges from ~8% (HWE) to ~85% (VRS), increasing consistently as scenarios become more restrictive and the system intensifies deployment on remaining non-forest sites. The higher forest capacity potential used reflects that, at equal onshore capital costs inside and outside forests, the system preferentially selects abundant southern forest sites for their proximity to demand centers, deploying a larger fraction of available forest capacity than of non-forest capacity in most scenarios [18,21]. CUL is the exception, where forest and non-forest capacity potential used converge at ~30%, because CUL's cultural landscape exclusions remove northern non-forest land while retaining southern forest areas where both land types offer comparable system value.

The gap in capacity potential used between forest and non-forest starts to decrease only under the most restrictive scenarios. From HWE through ECS, forest capacity potential used is equal to or exceeds that of non-forest, indicating that the system treats forest as its primary onshore resource. Both ECS and VRS show the widest configuration sensitivity, with forest capacity potential used spanning approximately 20–75% under ECS and 20–65% under VRS, reflecting that the few remaining forest sites become highly sensitive to solar and transmission assumptions. VRS is the only scenario where non-forest capacity potential used (~85%) substantially exceeds forest (~40%), indicating that the system shifts to non-forest and offshore resources only when forest options are effectively eliminated, consistent with the cost analysis finding that VRS incurs the highest system cost penalty (16–22%) precisely because this forced substitution requires capital-intensive offshore investment **(Fig. 2 & 3)**.

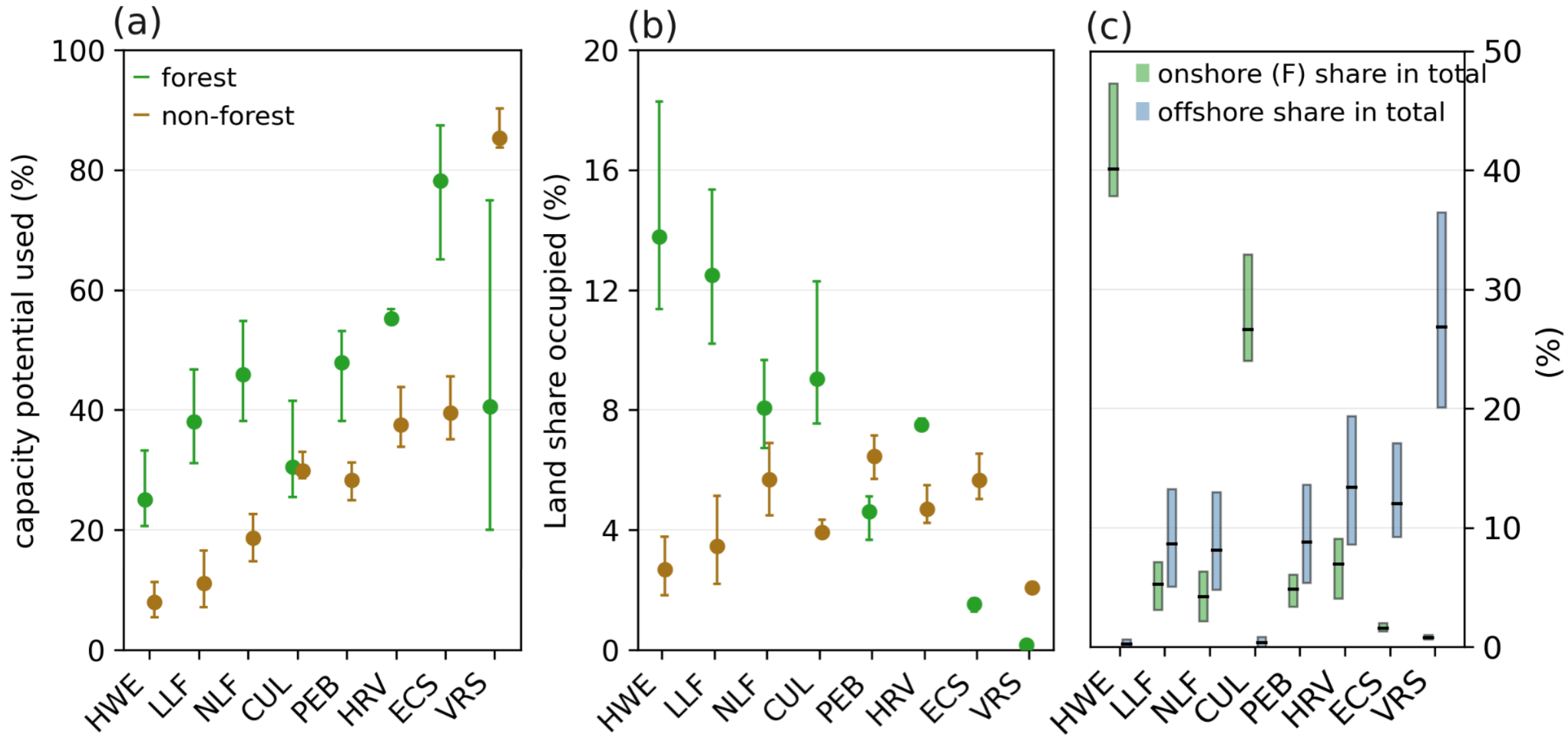

**Fig. 4.** Forest-specific deployment analysis for a 2050 net-zero national electricity system. (a) Capacity potential used: percentage of available (scenario-permitted) onshore wind capacity deployed in forest (green) and non-forest (brown) areas. Dots indicate mean values across system configurations **(Table 1)**; range bars show the minimum–maximum spread. (b) Land share occupied: percentage of Norway's total forest and non-forest area occupied by onshore wind deployment, using a capacity density of 2.4 MW/km². (c) Share of onshore wind capacity deployed in forests (green boxes, right y-axis) and offshore wind share of total national capacity (blue boxes, right y-axis). Box plots capture variation across all buffer distances and system configurations; black horizontal lines indicate medians.

Fig. 4b translates these capacity potential used rates into equivalent physical land impact, revealing that forest areas bear the dominant land footprint. Onshore wind deployment occupies 5–14% of Norway's total forest area across scenarios, while non-forest land occupation remains below ~6% in all cases. HWE and LLF produce the highest mean forest land occupation (13–14%) as the system deploys extensively across available forest areas. Under scenarios incorporating ecological and forest productivity criteria (ECS, VRS), forest land occupation drops below 2%, while non-forest occupation rises to 4–5%—demonstrating that stringent land-use preferences redirect the physical land burden from forests to non-forest areas at a system cost penalty of 8–22% relative to HWE (**Fig. 2)** [8,33].

The relationship between forest land occupation and its contribution to the national capacity mix reveals the system's dependence on forest resources **(Fig. 4c)**. Under HWE, ~14% forest land occupation yields an onshore forest share of approximately 40% of total national capacity — a dominant contribution that underscores how central forest deployment is to the cost-optimal system design. CUL achieves a similarly high onshore forest share (~28%) from ~9% forest land occupation, while requiring near-zero offshore capacity (~1%) — illustrating that when forest land is available, the system can deliver a net-zero electricity supply almost entirely through onshore resources, avoiding the capital-intensive offshore investment that drives cost increases under restrictive scenarios. Conversely, ECS and VRS show onshore forest shares below 2%, coinciding with offshore shares of 20–35%, suggesting that their land-use preferences effectively force the system into offshore dependence [43].

The inverse relationship between onshore forest share and offshore share across scenarios quantifies the substitution boundary between these two resource pools **(Fig. 4c**). Under HWE, ~40% forest share coexists with ~1% offshore share; under VRS, ~1% forest share requires ~35% offshore share. For intermediate scenarios (LLF, NLF, PEB, HRV), both metrics and their configuration-driven ranges remain relatively stable, suggesting that variation from buffer and solar assumptions is absorbed through adjustments in non-forest onshore capacity and solar rather than through shifts between forest wind and offshore. The gap between onshore forest share and offshore share within each scenario ranges from ~39 percentage points under HWE to approximately 18–20 percentage points of offshore dominance under VRS, with CUL showing the most striking profile: ~28% forest share alongside ~1% offshore, a 27 percentage point gap that reflects CUL's role as the scenario most favorable to an onshore-dominated system design. This progressive shift from forest-dominated to offshore-dominated capacity mixes reflects that each exclusion framework fundamentally reshapes not just how much onshore wind is built, but whether the system relies on distributed onshore resources or concentrated offshore infrastructure—a structural choice with implications for investment risk, transmission requirements, and regional techno-economic distribution.

### 3.4 System efficiency under land-use restrictions

Onshore wind curtailment rates range from 1–15% across the full parameter space of exclusion scenarios, buffer distances, and system configurations **(Fig. 5a)**. Two distinct curtailment patterns emerge, stratified primarily by solar deployment level: solar-constrained configurations (S13, S30) cluster around 4–6% curtailment (lower dashed line), while higher-solar-saturation configurations (S40, S+) mostly shift upward to 6–9% curtailment (upper dashed line). This 2–3 percentage-point separation persists across all scenarios and capacity levels, suggesting that the solar saturation level is a stronger determinant of onshore wind curtailment than exclusion criteria, buffer distance, or transmission assumptions. The likely mechanism is that increased solar penetration increasingly satisfies daytime demand near demand centers, compressing the temporal window in which onshore wind has dispatch value and pushing a larger share of wind output into periods of lower demand or grid congestion [17,63,65,66].

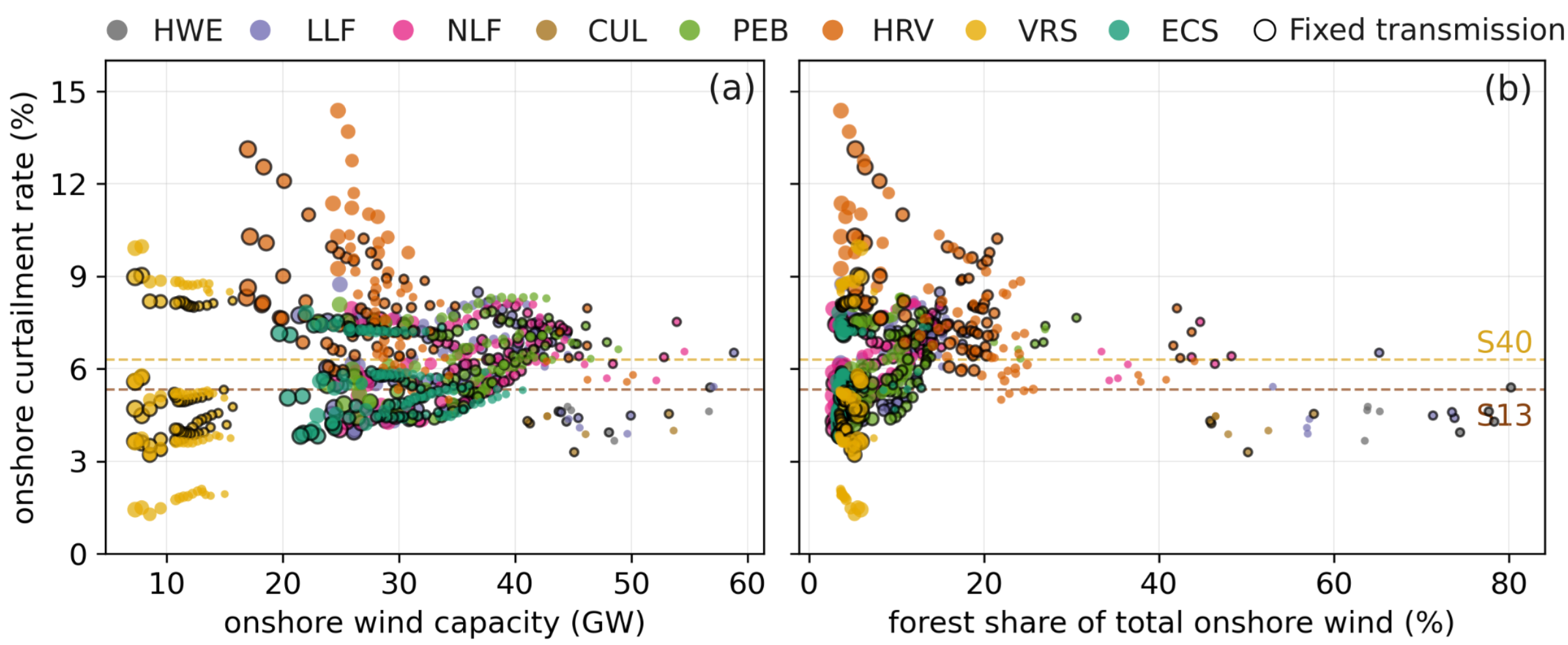

**Fig. 5.** System efficiency of onshore wind under land-use restrictions. Each dot represents a unique scenario-buffer-configuration combination across all scenarios (Table A1). Dot color indicates exclusion scenario; black border indicates fixed transmission; dot size is proportional to buffer distance (smallest = 0m, largest = 3000m). (a) Onshore wind curtailment rate versus total onshore wind capacity. (b) Onshore wind curtailment rate versus forest share of onshore wind capacity. Dashed horizontal lines indicate mean curtailment rates for solar-constrained (S13, brown) and moderate solar expansion (S40, gold) configurations. See **Fig. A1** for battery storage requirements across the same parameter space.

Scenarios with higher onshore capacity potential (30–57 GW under HWE, LLF, NLF) achieve comparable or lower curtailment rates than scenarios with restricted capacity potential (7–20 GW under VRS, HRV) (**Fig. 5a**). The likely explanation is that higher-capacity-potential scenarios distribute wind deployment across more counties, reducing spatial and temporal correlation of output. More restrictive scenarios lack this geographic diversification, concentrating deployment in fewer regions and increasing the likelihood of simultaneous generation peaks during high-wind events that the system cannot fully redistribute or absorb [65,66]. HRV exhibits this concentration effect most prominently, reaching 12–15% curtailment at moderate capacity levels (15–25 GW), the highest curtailment of any scenario, because its recreational area exclusions remove geographically dispersed sites while retaining spatially clustered locations (**Fig. 1**).

Buffer distance, encoded as dot size, shows a secondary effect within each scenario. As buffer increases, dots shift leftward (reduced capacity) with only marginal upward movement; curtailment increases slightly but not proportionally to the increasing buffer range (0-3000 m). This likely reflects that the spatial patterns of available capacity potential, which are predominantly shaped by each scenario's exclusion criteria, do not fundamentally change with buffer distance—the buffer reduces the available sites' area but likely preserves their relative spatial distribution. HRV is the notable exception, where larger dots climb more visibly upward, suggesting that the interaction between HRV's recreational exclusions and expanding buffer progressively concentrates deployment into fewer regions, amplifying co-generation effects that increase curtailment beyond what capacity reduction alone would predict. A broader system-level pattern is also visible: across all scenarios and configurations, the dot cloud trends slightly upward as total onshore capacity increases from ~20 GW to ~50 GW, with curtailment rising from approximately 4% to 7–8%. This suggests a weak but systematic relationship between total onshore wind volume and curtailment, indicating that at the highest deployment levels the system begins to encounter temporal saturation—even geographically distributed wind output increasingly coincides across regions, producing surpluses that Norway's hydropower fleet and interconnections cannot fully absorb.

**Fig. 5b** directly addresses whether deploying wind in forests creates system integration challenges. Forest share of total onshore wind capacity ranges from near zero (ECS, VRS) to ~70-80% (HWE, LLF), yet curtailment shows no systematic relationship with forest share. Points spanning 1–80% forest share occupy a broadly stable curtailment band of 3–8%, with even the highest forest shares (~60–80% under HWE and LLF) achieving curtailment rates of 4–7%—comparable to or lower than scenarios with minimal forest deployment. The vertical spread at any given forest share, approximately 3–4 percentage points, is driven primarily by solar configuration rather than by scenario exclusion criteria, illustrating that forest deployment level does not independently determine system integration performance. Our modelling results suggest that commonly cited concerns about wind deployment in forests, including reduced capacity factors due to canopy roughness [18,67], greater distance from demand centers, and weaker grid connections [62], do not materialize as system-level integration challenges in the Norwegian context. Forest sites selected by the system perform comparably to non-forest sites in terms of curtailment, likely because the spatial optimization inherently selects only those forest locations where resource quality and grid accessibility are sufficient for efficient dispatch.

Connecting with our previous findings, the 5–22% cost increases from HWE observed under restrictive scenarios are not primarily driven by worsening system efficiency. Curtailment rates do not substantially increase even as onshore capacity contracts by ~40 GW between HWE and VRS. This stability is partly attributable to Norway's large dispatchable hydropower fleet, which provides a significant buffer against curtailment by absorbing wind surpluses during high-production periods, and to existing import/export interconnections that enable cross-border balancing [6,16]. The cost increases are instead driven by the capital cost premium of substituting onshore wind with other technologies; in other words, restrictive scenarios cost more because they require more expensive generation assets, not because the system operates less efficiently with the assets it has.

### 3.5 Regional deployment profiles

The county-level analysis in Fig. 6 shows that forest exclusion preferences redistribute onshore wind deployment in scenario-specific spatial patterns that cannot be inferred from the national technology capacity mix alone **(Fig. 3)**.

Agder, the southernmost county (52% forest cover), illustrates how onshore wind deployment is concentrated in southern Norway under permissive forest scenarios. Under HWE and CUL, approximately 30–35% of Agder's total land area hosts onshore wind (red line at 0.70–0.75 on the normalized scale), contributing 15–20% of national capacity (blue line at 0.5–0.6). Agder's buffer sensitivity (green) is among the highest in the dataset, reaching 0.75–1 under permissive scenarios, corresponding to approximately 7.5–10 GW of capacity change between 0m and 700m setback distance, reflecting that its deployment of forest wind in a relatively compact geographic area makes it acutely vulnerable to setback regulation. Under strict scenarios (i.e., ECS, VRS), Agder's total area occupation contracts below 10% as ecological restrictions remove its forest wind sites, yet its national share declines less steeply, broadly because the overall onshore contribution to the national mix also falls under these scenarios **(Fig. 3)**.

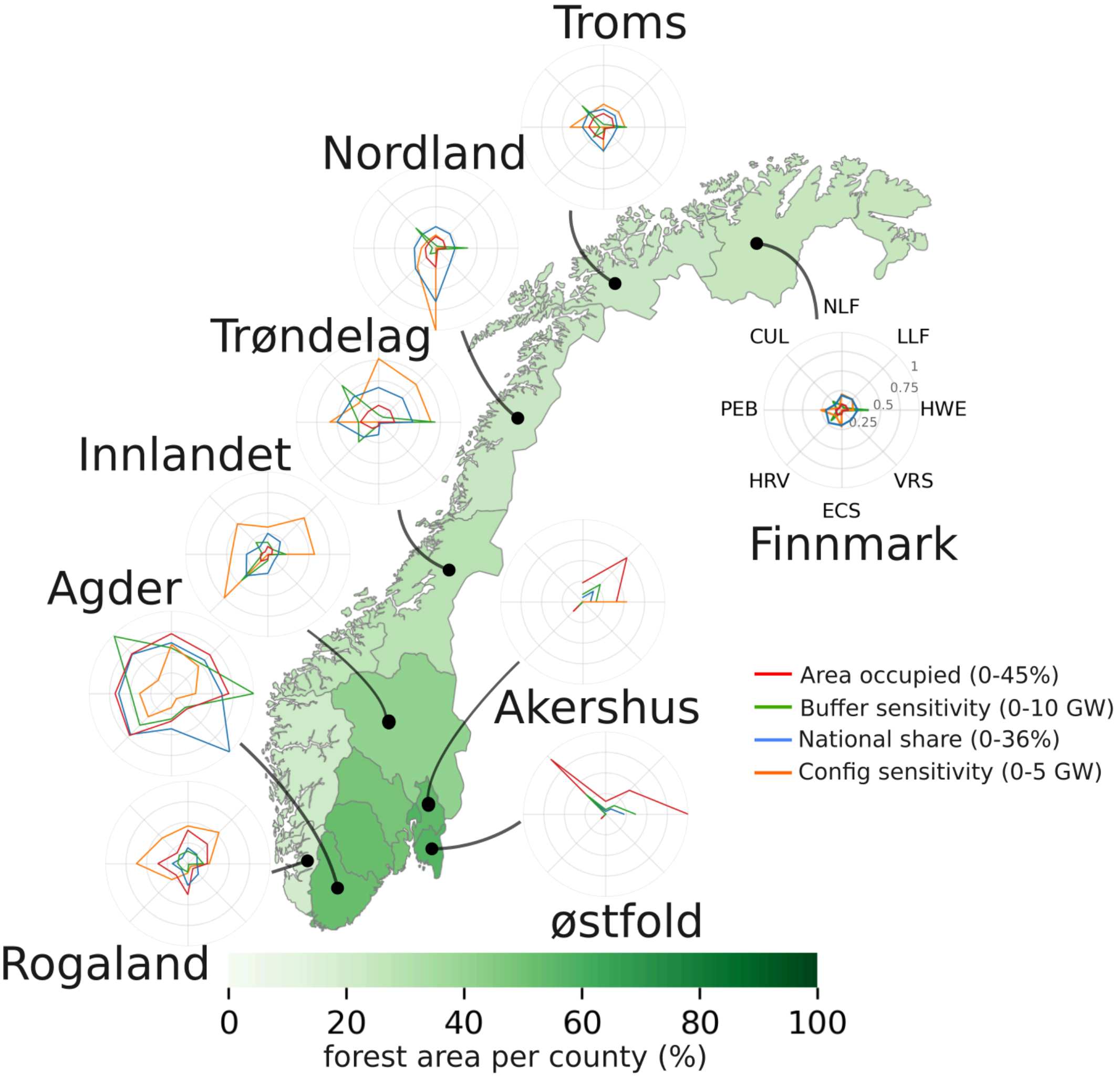

**Fig 6.** Regional deployment profiles across scenarios. The map shows county-level forest coverage (% of county area) derived from CORINE land cover data. Spider charts display four normalized deployment indicators for selected counties: area occupied (red; percentage of total county area used for onshore wind, range 0–45%), national share (blue; county's contribution to total national onshore wind capacity, 0–36%), buffer sensitivity (green; absolute capacity change between 0m and 700m buffer distance averaged across system configurations, 0–10 GW), and configuration sensitivity (orange; range of deployed capacity across solar and transmission assumptions, 0–5 GW). All indicators are min-max normalized to 0–1 across all counties and scenarios; radial gridlines indicate 0.25, 0.5, 0.75, and 1.0 of the observed range.

Østfold, the easternmost county and one of Norway's smallest (4,180 km², 57% forest cover), illustrates how land impact can scale inversely with county size. Under HWE, the share of county area occupied by onshore wind reaches the upper end of the observed range (45% of Østfold's area), yet its contribution to national onshore wind capacity remains below 5%. This disproportionate land burden relative to system contribution highlights a distributional equity concern: small, densely forested counties near population centers absorb significant local land-use impacts while contributing only marginally to the national electricity system [47,68]. Akershus, similarly with high forest coverage, shows a comparable pattern—elevated area occupation and configuration sensitivity under permissive scenarios but limited national share — reinforcing that the southern forested counties surrounding major demand centers collectively bear concentrated land-use impacts under forest-permissive exclusion frameworks.

Nordland and Troms present a contrasting northern profile. Their spider shapes remain compact across most scenarios — national share holds between 10–20% and area occupation between 10–15%. This stability reflects their lower percentage of forests in total land areas (green filled colour) and reliance on non-forest coastal wind sites unaffected by forest-based exclusion criteria. However, under ECS, national share and area occupation increase as the system redirects substantial capacity northward to compensate for the near-total exclusion of southern forest sites. Buffer and configuration sensitivities remain low for both counties, likely reflecting their large geographic extent, which dilutes the spatial impact of setback regulation, combined with limited solar potential at northern latitudes, which reduces sensitivity to solar deployment assumptions.

Finnmark displays a distinctive but subdued profile: ~2 GW deployment across most scenarios. Although its land is available across scenarios, Finnmark is rarely selected, as it is less economically attractive than southern and western sites or offshore alternatives. When the system does deploy onshore wind in the north, it primarily chooses Nordland and Troms. The only visible activity is moderate configuration sensitivity (orange) in some scenarios, showing Finnmark's role as a marginal region whose onshore resources are accessed only under specific combinations of exclusion and technology assumptions.

Innlandet, Norway's largest forested county (41% forest cover), is among the most sensitive counties to both land-use preferences and system configurations under comparatively less restrictive scenario frameworks. Under HRV, Innlandet shows the highest configuration sensitivity (orange) of any county-scenario combination — the orange line extends substantially outward, indicating that solar and transmission assumptions determine whether the system channels deployment into Innlandet's remaining forest sites or bypasses them entirely. Under CUL and HWE, Innlandet contributes upto 10% of national capacity with elevated buffer sensitivity, as its forest sites sit at the economic margin where setback regulation directly determines viability. This configuration sensitivity diminishes under the most restrictive scenarios (ECS, VRS), where most of Innlandet's sites are excluded regardless of technology assumptions.

Rogaland, despite low forest coverage (19%), shows the most pronounced configuration sensitivity (orange) spread across moderate restrictive scenarios, similar to Innlandet. Its spider chart expands substantially under permissive conditions, with the orange line reaching further outward than in most other counties, while contracting under restrictive conditions (ECS, VRS). This reflects that Rogaland operates at the intersection where solar deployment assumptions interact most strongly with land availability—under solar-constrained configurations, Rogaland's non-forest wind sites are fully utilised, while solar expansion reduces their dispatch value [63]. Changes in exclusion criteria and solar growth determine whether Rogaland develops intensive onshore capacity or substitutes capacity to other nearby regions through solar expansion.

The geographic pattern across all counties illustrates that the regional impact of exclusion policies is inherently scenario-dependent rather than following a uniform spatial gradient [40]. Under HWE, deployment concentrates in Agder, Trøndelag, and Østfold where forest wind resources and grid access converge; under VRS, Agder retains the highest national share as it holds the largest remaining available area under this scenario's exclusion criteria. No single policy parameter produces a homogeneous effect across all counties: buffer regulation disproportionately affects densely forested southern counties (Østfold, Agder, Trøndelag, Akershus) [55]; solar deployment assumptions most strongly influence counties at the economic margin of forest deployment (Innlandet, Rogaland); and exclusion criteria determine whether northern counties (Nordland, Troms) or offshore capacity absorb the displaced southern deployment. Each exclusion framework creates a distinct regional pattern of deployment burden—no single scenario minimizes impacts across all counties simultaneously.

## 4. Discussions

Our results indicate that the way forests are represented in energy system analysis can materially shape future energy system designs. Prior assessments have largely treated forests as a single category, either fully open to wind or excluded outright [16,17], yet forests differ widely in ecological value and are for the most part, managed production forest rather than intact habitat [30]. Grading forest exclusion by ecological value and coupling it to a spatially explicit net-zero energy system modelling allows the ecological value forgone per unit of wind deployed to be quantified, an accounting that studies identify as largely missing from wind siting assessment frameworks [13,35]. Our results then suggest that, at equal cost, the system treats forest as its preferred onshore resource, so that excluding forest carries a more pronounced cost penalty than analyses assuming costlier forest wind would imply (Fig. 2 & 3). Yet the marginal cost of forest protection declines as exclusions tighten, indicating that the ecologically graded scenarios can spare forests of comparatively high biodiversity value at modest system cost. In the Norwegian case, the system cost remains at around 5–10% across most scenarios relative to the high wind scenario (Fig. 2).

A recurring objection to siting wind in forests is that forested terrain makes a poorer host due to reduced capacity factors under canopy roughness, greater distance from demand, and weaker grid access [20,67]. Our results suggest these concerns need not translate into system-level penalties. Onshore curtailment stays within a narrow 3–8% band regardless of the forest share of deployment, and the forest sites the system selects perform comparably to non-forest ones, likely because the system retains only forest locations with adequate resource quality and grid access. Restricting forest wind does not erode this efficiency either. Curtailment remains stable even as onshore capacity contracts substantially, likely reflecting the large firm dispatchable hydropower resource and cross-border interconnections [6,16,65]. The added cost of protection therefore reflects substitution toward more capital-intensive offshore wind generation, not a less efficient system [17]. This findings reframe the debate over wind energy deployment in forests. Because avoiding forest deployment does not necessarily compromise overall system integration and efficiency, the case for protecting forests rests on their ecological and social value, not on technical grounds.

Land-use choices and technology adoption policies are often treated separately, yet our results suggest they cannot be [7,17]. The system value of onshore wind is not fixed but depends on other system-level policy decisions. For example, where onshore wind is abundant, added solar competes with it for daytime dispatch and compresses the temporal window in which onshore wind holds dispatch value [63,66] (Fig. 3). Where onshore deployment wind is scarce, the same solar saturation levels instead displace offshore capacity, so onshore wind retains its value due to technology diversification. Solar therefore does not displace wind uniformly but displaces whichever resource dominates the mix, forest wind under permissive scenarios and offshore wind under restrictive ones. This asymmetry shows that a land-use decision and a solar technology adoption,  which appear independent, may in fact interact, so the resulting system is difficult to infer from either choice alone. Storage follows the same logic, with its scale set mainly by solar saturation rather than by which forests are protected [65] (Fig. A1).

These scenario preferences are applied across spatially varied landscapes, and their regional consequences are correspondingly uneven. Our results indicate that no single ecological scenario minimises local impacts across all regions [29,35]. The burden a given scenario places on a given county depends on the interaction among its forest and ecological character, its geography, and the prevailing solar and transmission assumptions, so that a preference that eases pressure in one region can intensify it in another (Fig. 6). Under permissive scenarios, small and densely forested counties near demand centres bear a disproportionate share of local land-use impacts despite making only a modest contribution to national electricity supply. This spatial heterogeneity suggests that the regional distribution impacts under ecologically graded scenarios may warrant explicit attention in onshore wind-siting policy [40,69,70].

Meeting climate targets requires onshore wind to expand far faster than its recent pace, yet deployment has repeatedly stalled where it faces socio-ecological opposition, diverging from the least-cost path assumed by models [71]. This expansion is driven by the electrification of energy demand, and in economies such as Norway it also enables industrial diversification beyond oil and gas [16]. Our results point to a practical way forward: Although the system favors wind deployment in forests, protecting the most ecologically valuable forests remains compatible with achieving a net-zero electricity system. Opening all forests to wind energy deployment may unnecessarily sacrifice ecological value. Conversely, excluding all forests may increase reliance on more expensive technological options [17], reduce the spatial and technological diversification that net-zero systems depend on [72], and, by slowing decarbonisation, may itself contribute to irreplaceable ecological loss [32]. Ecologically graded exclusion offers a middle path, weighing high-biodiversity-value forest explicitly against land-intensive low-carbon generation so that forest carbon and onshore wind compete on comparable terms. In most scenarios, protecting higher-value forest entails only a modest increase in system costs of ~5–10%. Since the trade-off between climate mitigation through renewable deployment and forest conservation varies across regions and forest types [31], the framework can be extended across Europe and beyond, navigating the same trade-offs and decisions. By quantifying the ecological value forgone per unit of onshore wind, it allows a contested choice to be settled through informed public deliberation rather than left implicit within an energy-system optimisation.

## 5. Limitations and future work

Several limitations are worth considering when interpreting our results. The scenarios rest on specific ecological thresholds for ecological datasets, set through Jenks classification, environmental-impact-assessment conventions, and forestry expertise. Alternative thresholds may shift the numerical results while likely preserving the trade-off structure, so the figures are best read as illustrative of trade-off magnitudes within the ecologically graded scenarios rather than as objective predictions. Our approach nonetheless provides a means to test such sensitivities systematically as the quality and spatio-temporal resolution of ecological data improve. Onshore wind within and outside the forest is represented using identical techno-economic parameters, and incorporating site-specific differences in access or construction costs could further refine the estimated forest preference. The system is optimized for a single 2050 target year under perfect foresight and a single weather year, so it does not capture investment and weather uncertainty, path dependency, or interannual variability. Multi-year, pathway-based modelling is a natural extension [73].

In addition, the demand projections do not yet include deep-decarbonisation pathways, and hydrogen and synthetic-fuel loads could raise demand and tighten the more restrictive scenarios [45]. Our accounting of the ecological value forgone per unit of wind complements, rather than replaces, evidence that total electricity demand is the dominant driver of biodiversity pressure [33]. Future work could elicit exclusion thresholds through participatory processes with communities, developers, and conservation organisations to strengthen legitimacy [43], and could extend the ecologically graded approach to other forested systems facing similar competing climate strategies [31].

## 6. Conclusions

Onshore wind is increasingly sited in forests, emerging wind generation and forest carbon as competing climate solutions for the same land, alongside other ecological concerns. Representing forests as ecologically graded rather than as a single category illustrates that this tension need not be resolved by opening all forest or excluding it entirely. Our study shows that, at equal cost, the modelled net-zero system treats forest as its preferred onshore resource. Even so, the most ecologically valuable forest can be protected at a modest increase in system cost, so a graded approach can reconcile substantial wind expansion with the protection of high-value forest.

Several system-level findings underpin this conclusion. Commonly cited concerns about forest wind, including reduced capacity factors under canopy roughness, distance from demand, and weaker grid access, do not translate into system-level integration penalties, and the added cost of protection reflects substitution toward more capital-intensive generation rather than a less efficient system. The case for protecting a given forest therefore rests on its ecological and social value rather than on technical grounds. This substitution is not unlimited. As both forest and non-forest land become constrained, the marginal cost of exclusion rises, and the system's capacity to substitute reaches a threshold. These outcomes are shaped as much by other low-carbon technology adoption levels as by land-use preference, since solar deployment reshapes the value of onshore wind and displaces whichever resource dominates the generation mix, so that land-use and technology decisions cannot be made independently and their consequences may vary from region to region.

By quantifying the ecological value forgone for each unit of wind deployed, the ecologically graded approach makes explicit a choice that energy system modeling usually leaves implicit, opening it to public deliberation. For forest-rich countries pursuing rapid decarbonisation, this offers a transferable basis for reconciling wind expansion with forest protection through informed settlement rather than blanket rules.

**CRediT authorship contribution statement**

Muhammad Shahzad Javed: conceptualisation, methodology, data curation, formal analysis, visualisation, writing—original draft; Marianne Zeyringer: conceptualisation, methodology, visualisation, writing—review & editing; Johannes Rahlf: conceptualisation, data curation,

writing—review & editing; Oliver Moen Snoksrud: conceptualisation, data curation, writing—review & editing;

## Declaration of competing interest

The authors declare that they have no known competing financial interests or personal relationships that could have appeared to influence the work reported in this paper.

## Data availability

The highRES model version used in this work is openly available on GitHub: https://github.com/JavedMS/highRES-Norway/tree/highres_forestwork. All underlying datasets are openly available and provided with access links in Table A2. Processed data and scenario input files can be requested from the authors for further scenario design–related queries.

## Acknowledgments

This research received no specific grant from any funding agency in the public, commercial, or not-for-profit sectors.

# Appendix:

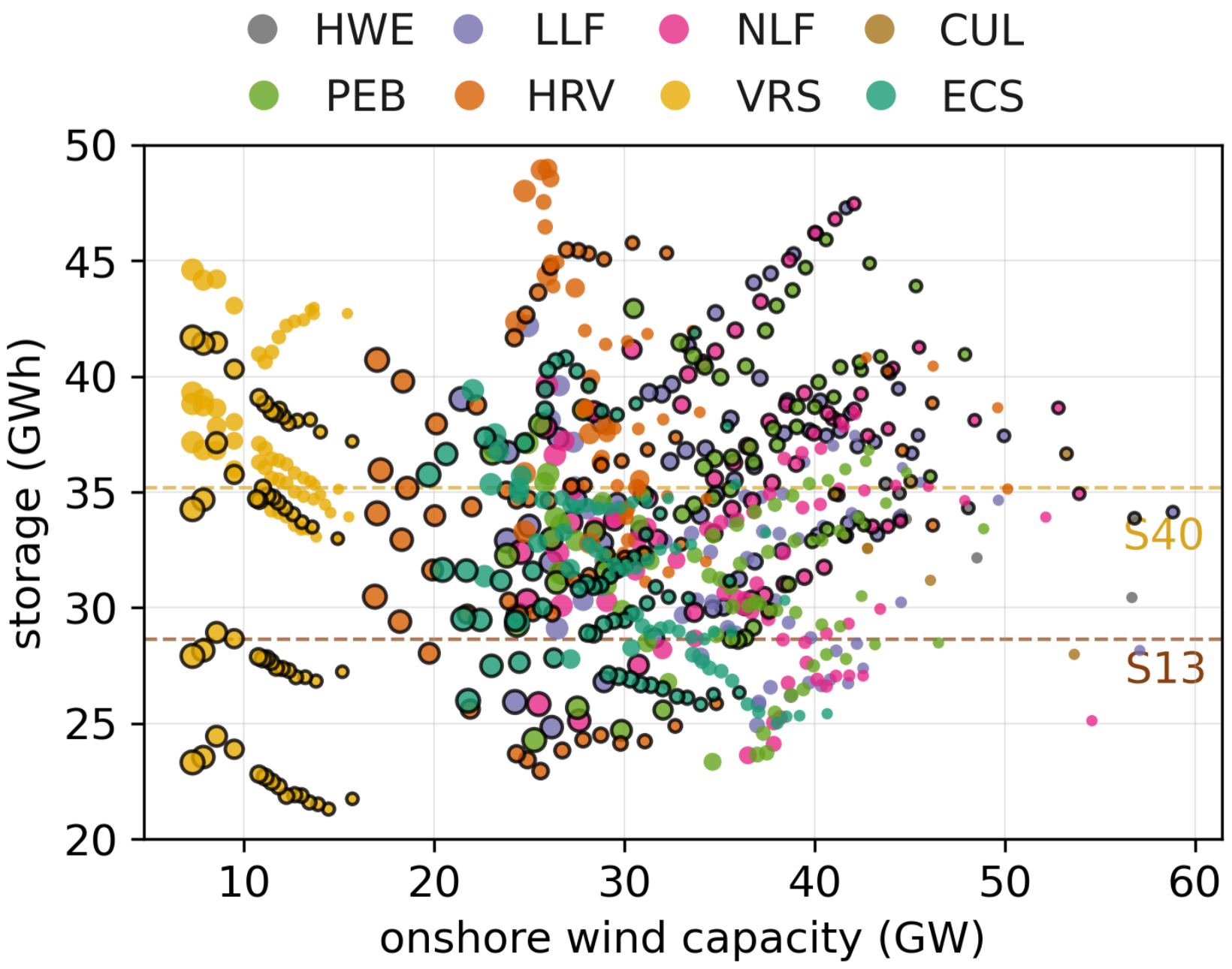

**Fig. A1.** Battery storage requirements for 2050 net-zero electricity system across scenarios and system configurations. Each dot represents a unique scenario-buffer-configuration combination across all scenarios (Table A1). Dot color indicates exclusion scenario; black border indicates fixed transmission; dot size is proportional to buffer distance (smallest = 0m, largest = 3000m). Battery storage capacity (GWh) is shown against total onshore wind capacity. Dashed horizontal lines indicate mean storage requirements for the solar-constrained (S13, brown) and moderate solar expansion (S40, gold) configurations.

**Table A1**. Ecological and forest exclusion criteria for all scenarios. Each column lists the criteria excluded from onshore wind development under a scenario, ordered from the most restrictive (VRS) to the most permissive (HWE). Scenarios exclude only the protected-area categories and ecological, forest, recreation, and reindeer criteria relevant to their priorities, so that a criterion left blank is available for development under that scenario. Thresholds for the continuous surfaces follow the classification described in Section 2.1, with data sources given in Table A2.

| | **VRS** | **ECS** | **NLF** | **LLF** | **PEB** | **HRV** | **CUL** | **HWE** |
|---|---|---|---|---|---|---|---|---|
| **protected areas** | Ia, Ib, II, III, IV,V | Ia, Ib, II, IV | Ib, II, III | Ib, II, III | Ib, II, | Ia, Ib, II, III, V | Ia, Ib, V | Ib, II |
| **Redlist species with no buffer** | Yes | Yes | — | — | — | — | — | — |
| **Peat lands exclusion** | Yes | — | — | — | — | — | — | — |
| **Scenario specific** | Forest<1940 | Forest<1940 | Forests ages>105 | Forests ages>105 | Forest<1940 | Forests ages>105 | — | — |
| | Naturalness>1 | Naturalness>3 | — | — | Naturalness>5 | — | Naturalness>3 | — |
| | Recreational areas>1with 400 m buffer | — | — | — | Recreational areas>=5 with 50 m buffer | Recreational areas>1 with 400 m buffer | — | — |
| | Birds high | Birds high | — | — | Birds low | — | — | — |
| | All productive forests | — | All productive forests | Exclude productive forests> 20 & <60 | Productive forests>11 | Productive forests>11 | — | — |
| | Biomass volumes >59 | — | Biomass volumes >59 | — | Exclude biomass volumes >105 | Exclude biomass volumes >105 | — | — |
| | Reindeer grazing areas (summer & fall) & migratory routes with 0 m buffer | — | — | — | — | — | Reindeer grazing areas (all) & migratory routes with 0 m buffer | — |
| | non-flying mammals>5.37 | non-flying mammals>5.37 | — | — | non-flying mammals>7.92 | — | — | — |

*Recreational areas are categorized based on the number of visitors, which ranges from 1 to 5, where 1 means very few and 5 means a very high number of people visit.

**Naturalness data ranges from one to seven. The higher the value, the higher the degree of naturalness. It is based on predictions for share of dead wood, variation of dead wood and development stage of the forest.

***Productive forest data ranges from 0 to 23, with 23 indicating the most productive forests.

****The biomass volume of forests ranges from zero to 826.

*****A minimum 400-meter buffer is recommended to limit wind turbine noise below 50db

**Table A2.** Geodata

| Criteria | Dataset/source |
|---|---|
| Bird and mammal species richness | https://hdl.handle.net/11250/5173311 |
| Red-listed species | https://kartkatalog.geonorge.no/metadata/roedlistearter/00b0ab63-de0b-4f12-ac40-655de5163b8d |
| Naturalness (pre-1940 no clear-cut; natural-forest probability; naturalness gradient 1–7) | https://kartkatalog.geonorge.no/metadata/naturskog-v1/a0062ac4-8ee0-408f-9373-4c8b8c3088d8 |
| Forest productivity (index:0-23), standing biomass and age | https://www.nibio.no/tema/skog/kart-over-skogressurser/skogressurskart-sr16 ; https://link.springer.com/article/10.1186/s40663-021-00338-4 |
| Peatlands, water bodies, urban areas, forest / non-forest | https://land.copernicus.eu/en/products/corine-land-cover/clc2018?tab=download |
| Protected areas (IUCN categories Ia, Ib, II, IV, V) | World Database on Protected Areas (WDPA), https://www.protectedplanet.net/en/thematic-areas/wdpa?tab=WDPA |
| Recreational areas | https://kartkatalog.geonorge.no/metadata/friluftslivsomraader-kartlagte/91e31bb7-356f-4478-bcba-d5c2de6e91bc |
| Reindeer seasonal grazing & migratory routes | https://kartkatalog.geonorge.no/metadata/reindrift-aarstidsbeite-vinterbeite-wms/cbe491fd-59cd-496a-bcd6-4e074de68d50; https://kartkatalog.geonorge.no/metadata/reindrift-aarstidsbeite-hoestvinterbeite-wfs/e19e7e0f-07a8-4636-b3f6-4ba560b1136d?search=Reindrift%20-%20%C3%85rstidsbeite; https://kartkatalog.geonorge.no/metadata/reindrift-aarstidsbeite-sommerbeite-wfs/31c7b0c7-0e76-45e5-8f52-3ec50d18fa39?search=Reindrift%20-%20%C3%85rstidsbeite; https://kartkatalog.geonorge.no/metadata/reindrift-aarstidsbeite-vrbeite-wms/b9cb06f7-0a87-4a55-8399-9c8118705b63?search=Reindrift%20-%20%C3%85rstidsbeite; https://kartkatalog.geonorge.no/metadata/reindrift-flyttlei/f9c1e228-892f-4f1a-9e4e-b6d6149f373c?search=Reindrift; |